\documentclass[12pt]{article}
\usepackage{epsf}
\usepackage{graphicx}
\usepackage{array,dcolumn}      
\usepackage{amssymb}
\newcommand{\be}{\begin{equation}}
\newcommand{\ee}{\end{equation}}
\newcommand{\bea}{\begin{eqnarray}}
\newcommand{\eea}{\end{eqnarray}}
\newcommand{\baa}{\begin{eqnarray*}}
\newcommand{\eaa}{\end{eqnarray*}}
\newcommand{\bary}{\begin{array}}
\newcommand{\eary}{\end{array}}
\newcommand{\bit}{\begin{itemize}}
\newcommand{\eit}{\end{itemize}}

\begin{document}
\title{\bf Virtual bilepton effects in polarized M\o ller scattering}
\author{B. Meirose and A. J. Ramalho  \\
Instituto de F\'{\i}sica \\
Universidade Federal do Rio de Janeiro \\
Caixa Postal 68528, 21945-970 Rio de Janeiro RJ, Brazil}

\maketitle
{\bf Abstract:}
We investigate the indirect effects of heavy vector bileptons
being exchanged in polarized M\o ller scattering, at the next
generation of linear colliders. Considering both longitudinal and
transverse beam polarization, and accounting for initial state
radiation, beamstrahlung and beam energy spread, we discuss how
angular distributions and asymmetries can be used to detect clear
signals of virtual bileptons, and the possibility of
distinguishing theoretical models that incorporate these exotic
particles. We then estimate $95\%$ C. L. bounds on the mass of
these vector bileptons and on their couplings to electrons.


\newpage
\section{Introduction}
 Several schemes have been put forward to address important
questions left unanswered in the standard model. There are models
in which some of the standard model fundamental particles, such as
quarks and leptons, are thought to be in fact composite. Another
approach is to extend the symmetry by enlarging the local gauge
group. In an effort to incorporate gravity, more exotic proposals
have been made, such as those which depend on the existence of
extra spatial dimensions. Since the standard model has passed
severe experimental tests at the $e^+e^-$ and hadron colliders,
showing its validity up to the electroweak scale, one expects that
it could be seen as an effective theory, valid up to some large
mass scale $\Lambda$. Most theoretical arguments point to
$\Lambda$ $\buildrel > \over \sim$ $1$ $TeV$. Extended electroweak
models predict the existence of new particles and interactions. An
interesting class of exotic particles known as bileptons
\cite{CUYDAV} is present in several of these models, such as
left-right symmetric models, technicolor and theories of grand
unification. The bileptons in which we are interested are vector
bosons which couple to standard model leptons and carry two units
of lepton number. In particular, heavy gauge bileptons may be
found in models where the standard $SU(2)_L \times U(1)$ group is
embedded in a larger gauge group. This is the case of some of the
$331$ gauge models \cite{FRAPLE}. In this paper we propose to look
for indirect signals of doubly-charged vector bileptons in M\o
ller scattering, with polarized beams, and at the new linear
collider energies. The next generation of $e^+e^-$ colliders will
be designed to operate at center-of-mass energies ranging from the
$Z^0$ mass to a few TeV, with very high luminosities. Electron and
positron beam polarizations are expected to be available at these
facilities, providing experimentalists with powerful tools to
carry out precision tests of the standard model and to explore new
physical phenomena. Currently, the best lower bound on the vector bilepton
mass is $M_Y > 850 \, GeV$, a result which was established from
muonium-antimuonium conversion \cite{WILL}. Another useful lower
bound $M_Y > 740 \, GeV$ has been derived \cite{TULL} from current
experimental limits on fermion pair production at LEP and
lepton-flavor violating charged lepton decays. While less
stringent, this limit does not depend on the assumption that the
bilepton coupling is flavor-diagonal.

\par
Studies of bilepton effects in M\o ller scattering have been made
before \cite{FRIZ}, mostly in the context of SU(15) grand unified
theories. In this paper we work in the framework of the $331$
models, concentrating on the minimal version, and extend the
previous studies by taking into account (i) important beam effects
such as initial-state radiation, beamstrahlung and beam energy
spread; (ii) longitudinal and transverse polarization of the
colliding beams, which are expected to be available in the next
generation of linear colliders; (iii) Gaussian smearing of the
four-momenta of the final-state leptons, simulating the
uncertainties in the energy measurements in the electromagnetic
calorimeters. All these items are significant for a realistic
comparison with experimental data. We analyze several
distributions and asymmetries in M\o ller scattering at the Next Linear Collider (NLC)
energies, with the purpose of searching for indirect signals of
gauge bileptons. The predictions of an $SU(15)$ GUT model
\cite{SUFI} for these observables are also shown for comparison.
From the angular distribution of final-state electrons we
establish bounds on the couplings and mass of these bileptons at
the $95\%$ confidence level. If evidence of the existence of these
vector bileptons is found in M\o ller scattering, one can also
verify the validity of relations between the masses of the vector
bileptons and new neutral gauge bosons, which are connected to the
Higgs structure of the $331$ models. In section II we give a brief
review of the $331$ models \cite{PPT}, outlining only those
features that are most relevant to our analysis. Section III
describes in detail the numerical simulation of the M\o ller
events and the corresponding analysis of the results. $95\%$
confidence level limits are established in section IV. Our
conclusions are summarized in section V.

\section{Review of the $331$ models}
\par
The $331$ models are based on the $SU(3)_C \otimes SU(3)_L \otimes
U(1)_X$ gauge group and predict new physics at the TeV energy
scale. They arrange the ordinary leptons in $SU(3)_L$
antitriplets, two generations of quarks in triplets and a third
generation in an antitriplet. Anomaly cancelation, essential to a
gauge theory, takes place not within each generation but when all
three families of quarks and leptons are taken together. The fact
that one quark generation transforms differently from the other
two is essential in these models, since the number of triplets
must equal the number of antitriplets to ensure that the models
remain anomaly-free. This implies that the number of generations
is divisible by the number of colors. Besides standard gauge
bosons, the models predict a new neutral gauge boson $Z^{\prime}$
and four vector bileptons $Y^{\pm}$ and $Y^{\pm \pm}$. These gauge
bosons are expected to be considerably heavier than the standard
gauge bosons. In addition to the ordinary quarks, the class of
models we are considering here contains three new heavy quarks,
with exotic electric charges $5/3$ and $-4/3$. However it is
possible to build $331$ models in which these exotic quarks are
not present \cite{PONCE}, and as a consequence, neither are
doubly-charged gauge bileptons. Following the notation of ref.
\cite{FRIZ}, the coupling of the doubly-charged vector bileptons
$Y^{++}$ to electrons and positrons is given by the interaction
Lagrangian
\begin{equation}
L_{int} = -{g_{3l} \over \sqrt{2}}Y_{\mu}^{++}e^TC \gamma^\mu
\gamma_5 e -{g_{3l} \over \sqrt{2}}Y_{\mu}^{--}\bar{e}\gamma^\mu
\gamma_5 C {\bar{e}}^T \hskip 0.1cm .
\end{equation}
For the minimal $331$ model, symmetry breaking can be accomplished
by three complex $SU(3)_L$ triplets and a complex sextet, allowing
these nonstandard gauge bosons to acquire plausible masses. In
this case, the mass $M_{Z^\prime}$ of the neutral vector boson
$Z^\prime$ and the mass $M_Y$ of the doubly-charged bilepton
$Y^{++}$ are related by
\begin{equation}
{M_Y \over M_{Z^\prime}} = {{\sqrt{3(1-4\sin^2\theta_W)} \over
2\cos \theta_W }}
\end{equation}
Alternative Higgs structures are possible in different $331$
models, but then the relation above no longer holds.

\section{Simulation and analysis}
\par
One substantial source of beam energy degradation is the
initial-state radiation (ISR). This is a QED effect, and
corresponds to the emission of photons  by the incoming electrons
and positrons. For a considerable fraction of events, this
bremsstrahlung emission lowers the effective center-of-mass energy
available for the hard scattering process. To account for ISR, we
used the structure function approach discussed in \cite{SKR}. In
order to achieve high luminosities at the new linear colliders,
the beams must have very small transverse dimensions. The
particles in a colliding bunch suffer considerable transverse
acceleration due to the collective electromagnetic fields produced
by the particles in the opposite colliding bunches, which gives
rise to the emission of synchrotron radiation, the so-called
beamstrahlung. The effective energy available for the reaction is
then smaller than the nominal value. The average energy loss of a
positron or electron by beamstrahlung depends on the design
parameters of the accelerator. For some designs of a $1 \, TeV$ NLC,
for instance, the colliding beams may lose about $13\%$ of the
nominal energy from beamstrahlung emissions. This loss may reach
$31\%$ at a $3 \, TeV$ CLIC \cite{CLIC}. To obtain the beamstrahlung
spectrum, we followed the approach of ref. \cite{PESK}, which is
based on the Yokoya-Chen approximate evolution equation for
beamstrahlung \cite{YCHEN}. In this paper, the calculations of the
beamstrahlung energy spectra were carried out starting from the
energy-dependent sets of NLC design parameters \cite{BEAM}. The
spectra corresponding to ISR and beamstrahlung emissions were
convoluted, and the resulting distribution was used to compute all
the required differential cross sections. We also considered a
possible beam energy spread, which was taken to be Gaussian
distributed, with a width of $1\%$ of the nominal beam energy.
\par
In addition to the standard Feynman diagrams for M\o ller
scattering, in $331$ models the process also proceeds via an
s-channel exchange of a doubly-charged bilepton, and a t-channel
exchange of a $Z^\prime$ boson as well. By ignoring beam effects,
neglecting the $Z^\prime$ exchanges and considering only
unpolarized beams, we verified that our numerical calculations
agree with the trace calculation of ref. \cite{FRIZ}. Likewise, we
cross-checked our calculations with those of ref. \cite{OLS},
which were carried out in the framework of the standard model, but
with arbitrary beam polarization.

 The differential cross sections were calculated
with Monte Carlo techniques, with the simulated events selected
according to the following set of cuts : (i) the final-state
electrons and positrons were required to be produced within the
angular range $\vert cos\theta_i \vert < 0.95$, where $\theta_i$
stands for the polar angle of the final-state lepton with respect
to the direction of the incoming electron beam; (ii) all events in
which the acollinearity angle $\zeta$ of the final-state $e^+-e^-$
three-momenta did not pass the cut $\zeta < 10^\circ$ were
rejected; (iii) the ratio of the effective center-of-mass energy
to the nominal center-of-mass energy for any acceptable event was
required to be greater than $0.9$ . Our simulations were made in
the context of the NLC program \cite{NLC}, considering a maximum
center-of-mass energy of $1$ $TeV$, and several years of
operation, so as to accumulate an integrated luminosity of $500$
$fb^{-1}$. In order to simulate the finite resolution of the NLC
electromagnetic calorimeters, we Gaussian-smeared the four-momenta
of the produced electrons and positrons \cite{SMEAR}. The energies
of the final-state leptons were distributed as a Gaussian with
half-width $\Delta E$ of the form $\Delta E/E = 10\%/\sqrt{E}
\oplus 1\%$, whereas the directions of the lepton three-momenta
were smeared in a cone around the corresponding original
directions, whose half-angle is a Gaussian with half-width equal
to $10$ $mrad$.

\par

     Beam polarization will play a useful role at the new linear
colliders \cite{MOOR}. By using polarized lepton beams, one can effectively
reduce backgrounds and increase the sensitivity of spin-dependent
observables to potential new physics. In our calculations we
worked with the electron beam projection operators in the extreme
relativistic regime,
$${\cal P}(p_{a,b})  = \lim_{m \rightarrow 0} \; \frac{1}{2} (\not\!{p}_{a,b}
+ m)(1+{\gamma}_5 \not\!{n}_{a,b}) \rightarrow$$
$$\frac{1}{2} (1 + P_L^{a,b} \gamma_5)\not\!{p}_{a,b} + \frac{1}{2} P_T^{a,b}
\gamma_5 (cos\phi_{a,b} \not\!{n}_1 + sin\phi_{a,b} \not\!{n}_2)
\not\!{p}_{a,b} \, ,$$ where $n_{a,b}^\mu$ represent the spin
vectors and $P_L^{a,b} (P_T^{a,b})$ stand for the longitudinal
(transverse) polarizations of the incoming electron beams, whose
four-momenta for a nominal center-of-mass energy $\sqrt{s}$ are
given by $p_a^{\mu} = (\frac{\sqrt{s}}{2},0,0,\frac{\sqrt{s}}{2})$
and $p_b^{\mu} = (\frac{\sqrt{s}}{2},0,0,-\frac{\sqrt{s}}{2})$.
For the numerical calculations dealing with transverse
polarization of the electron beams, the azimuthal angles of the
transverse polarization vectors were taken to be $\phi_a = \phi_b
= 0$, and the purely spatial vectors $n_1^\mu = (0,1,0,0)$ and
$n_2^\mu =(0,0,1,0)$.

In order to discuss how to probe the presence of gauge bileptons
in M\o ller scattering, we first considered longitudinally
polarized electron beams. For the longitudinal polarizations of
the beams we considered $P_L^a = -P_L^b = 0.9$, with an
uncertainty given by $\Delta P_L/P_L = 0.5\%$. Figure 1 shows the
dependence of the total cross section on the center-of-mass
energy, for an input bilepton mass $M_Y = 1.2 \, TeV$. Only for
higher energy values is the total cross section significantly
altered by the exchange of a vector bilepton. The angular
distribution $d\sigma/d(cos\theta)$ of the final-state electrons
is, however, more sensitive to the presence of such a particle.
This is displayed in Fig. 2, where the angular distribution is
plotted both for $\sqrt{s} = 500 GeV$ and $\sqrt{s} = 1 \, TeV$, with
the corresponding curves for the standard model shown for
comparison. The M\o ller scattering angular distribution in the
minimal $331$ model differs from that of the standard model for
most of the angular range, the more so for a center-of-mass energy
$\sqrt{s} = 1 \, TeV$. These deviations from the standard model
predictions can be used to establish bounds on the mass of the
gauge bilepton, and its coupling to electrons.  The symmetric
shape of the angular distribution suggests that the integrated
forward-backward asymmetry $A_{FB}$ should be small. This is
indeed the case, as shown in Fig. 3, where $A_{FB}$ is plotted for
several input values of $M_Y$, at an energy $\sqrt{s} = 1 \, TeV$,
and the one-standard-deviation error bars represent only the
statistical errors. Next we analyze the discovery potential of
spin asymmetries.

Starting from the polarization-dependent angular distributions, one can
compute the following asymmetries:
\begin{equation}
A_1(cos\theta) = \frac{d\sigma(-\vert P_L^a \vert, -\vert P_L^b \vert)
+
                    d\sigma(-\vert P_L^a \vert, \vert P_L^b \vert)  -
                    d\sigma(\vert P_L^a \vert, -\vert P_L^b \vert)  -
                    d\sigma(\vert P_L^a \vert, \vert P_L^b \vert)}
                   {d\sigma(-\vert P_L^a \vert, -\vert P_L^b \vert) +
                    d\sigma(-\vert P_L^a \vert, \vert P_L^b \vert)  +
                    d\sigma(\vert P_L^a \vert, -\vert P_L^b \vert)  +
                    d\sigma(\vert P_L^a \vert, \vert P_L^b \vert)  }
\end{equation}


\begin{equation}
A_2(cos\theta) = \frac{d\sigma(-\vert P_L^a \vert, -\vert P_L^b \vert)
- d\sigma(\vert P_L^a \vert, \vert P_L^b \vert)} {d\sigma(-\vert P_L^a
\vert, -\vert P_L^b \vert) + d\sigma(\vert P_L^a \vert, \vert P_L^b
\vert)}
\end{equation}


\begin{equation}
A_3(cos\theta) = \frac{d\sigma(-\vert P_L^a \vert, \vert P_L^b \vert)
- d\sigma(0,0) }{d\sigma(-\vert P_L^a \vert, \vert P_L^b \vert)  +
d\sigma(0,0) }
\end{equation}

\par
    In the limit $\vert P_L^a \vert = \vert P_L^b \vert = 1$, $A_1(cos\theta)$
and $A_2(cos\theta)$ reduce to the familiar parity violating M\o
ller asymmetries \cite{CZAR}
\begin{equation}
A_{LR}^{(1)} = {d\sigma_{LL}+d\sigma_{LR} -
d\sigma_{RL}-d\sigma_{RR} \over
d\sigma_{LL}+d\sigma_{LR}+d\sigma_{RL}+ d\sigma_{RR}} \nonumber
\end{equation}
and
\begin{equation}
A_{LR}^{(2)} = {d\sigma_{LL}-d\sigma_{RR} \over
d\sigma_{LL}+d\sigma_{RR}} \nonumber
\end{equation}
respectively.

\par
    The behavior of spin asymmetry $A_1$ as a function of $cos \theta$ is
depicted in Fig. 4, for a bilepton mass $M_Y = 1.2 \, TeV$. At
$\sqrt{s} = 500 GeV$ the deviation from the standard model values
is small. This asymmetry becomes more sensitive to the presence of
a vector bilepton at an energy of $1 \, TeV$, and its angular dependence
distinguishes the two nonstandard models . Statistical errors
for the spin asymmetries discussed in this section are rather
small, and most of the effects of the systematic errors are
expected to cancel out in these asymmetries. Fig. 5 shows that a
similar pattern holds for $A_2$, which is represented as a
function of $cos \theta$ for both $500 \, GeV$ and $1 \, TeV$. At this
latter energy, however, $A_2$ does not lead to a clear distinction
between the minimal $331$ model and the $SU(15)$ GUT model.
Asymmetry $A_3$ concerns the difference between a polarized
angular distribution and its unpolarized counterpart, and is
displayed in Fig. 6. Unlike $A_1$ and $A_2$, which take values of
the order of $5\%$, $A_3$ may become quite large, reaching a
maximum magnitude of about $63\%$ for the minimal $331$ model at
$\sqrt{s} = 1 \, TeV$. It can
be useful to discriminate a model with vector bileptons from the
standard model, even at a $500 \, GeV$ linear collider. At a
center-of-mass energy of $1 \, TeV$, the difference between the
prediction of the $331$ minimal model for  $A_3$ and the standard
values is even more striking.

\par
     We assume that in the next generation of linear colliders, transverse
polarization of electron beams will be available as an extra tool
to search for the new physics. This could be achieved by means of
spin rotators, which convert longitudinal into transverse
polarization. It is not immediately clear whether the use of
transversely polarized electron beams may add any important
information, which could not be extracted from longitudinally
polarized M\o ller scattering. In fact, standard M\o ller
scattering is known not to be particularly sensitive to transverse
beam polarization \cite{OLS}, but the presence of an s-channel
vector bilepton might modify this picture. Effects of transverse
beam polarization in the nonstandard M\o ller scattering under
discussion only materialize if both beams are transversely
polarized. In order to search for possible advantages of
transverse beam polarization for the problem at hand, we examined
in detail the behavior of the following differential azimuthal
asymmetry:

\begin{equation}
A_{1T}(cos\theta) = \frac{ \int_{(+)} d\phi \, {d^2\sigma \over
d(cos\theta)d\phi} - \int_{(-)} d\phi \, {d^2\sigma \over
d(cos\theta)d\phi}} { \int_{(+)} d\phi \, {d^2\sigma \over
d(cos\theta)d\phi} + \int_{(-)} d\phi \, {d^2\sigma \over
d(cos\theta)d\phi}} \qquad ,
\end{equation}
where the subscript $+(-)$ indicates that the integration over the
azimuthal angle $\phi$ is to be carried out over the region of
phase space where $cos2\phi$ is positive (negative). As far as
$A_{1T}(cos\theta)$ is concerned, and considering an integrated
luminosity of $500 \, fb^{-1}$, we found that it would be
difficult to separate the signal from the standard model
background at center-of-mass energies around $500 GeV$, even if
$M_Y$ is only moderately large. At $\sqrt{s} = 1 \, TeV$, however, it
is feasible to detect the effects of a vector bilepton, as long as
its mass is not much larger than the center-of-mass energy. This
is illustrated in Fig. 7, where $A_{1T}(cos\theta)$ is plotted
both for an input mass $M_Y = 800 GeV$ and and for $M_Y = 1.2
\, TeV$, along with the standard model expectation. We also
investigated the integrated version of the asymmetry above,

\begin{equation}
A_{2T} = \frac{ \int_{(+)} d(cos\theta) d\phi \, {d^2\sigma \over
d(cos\theta)d\phi} - \int_{(-)} d(cos\theta)d\phi \, {d^2\sigma
\over d(cos\theta)d\phi}}{ \int_{(+)} d(cos\theta)d\phi \,
{d^2\sigma \over d(cos\theta)d\phi} + \int_{(-)} d(cos\theta)d\phi
\, {d^2\sigma \over d(cos\theta)d\phi}} \qquad ,
\end{equation}
where the integrations are consistent with the cuts specified in
section III. The mass dependence of $A_{2T}$ for $\sqrt{s} = 500
GeV$ and $\sqrt{s} = 1 \, TeV$ is presented in Fig. 8. $A_{2T}$ is
found to be sizable over a fairly wide range
around the bilepton resonance. We checked that the shape of the
curve representing the mass dependence of this asymmetry becomes
wider as $g_{3l}$ increases, while the position of the
corresponding minimum remains essentially the same, for a fixed
center-of-mass energy.

\section{Bounds}
A $\chi^2$ test was applied to estimate discovery limits for a
vector bilepton in M\o ller scattering. Considering only
longitudinal beam polarization, we compared the angular
distribution $d\sigma/d(cos\theta)$ of the final-state electrons, modified by
the presence of a vector bilepton, with the corresponding standard
model distribution. Assuming that the experimental data will be
well described by the standard model predictions, we defined a
two-parameter $\chi^2$ estimator
\begin{equation}
\chi^2(g_{3l},M_Y) = \sum_{i=1}^{N_b} {\biggl( {N_i^{SM}- N_i
\over \Delta N_i^{SM}}\biggr)^2}
\end{equation}
where $N_i^{SM}$ is the number of standard model events detected
in the $i^{th}$ bin, $N_i$ is the number of events in the $i^{th}$
bin as predicted by the model with bileptons, and $\Delta N_i^{SM}
= \sqrt{(\sqrt {N_i^{SM}})^2 + (N_i^{SM}\epsilon)^2}$ the
corresponding total error, which combines in quadrature the
Poisson-distributed statistical error with the systematic error.
For the latter we assumed a conservative value $\epsilon = 5\%$
for each measurement. The angular range $\vert cos\theta \vert <
0.95$ was divided into $N_b= 20$ equal-width bins. The coupling
$g_{3l}$ of a vector bilepton to electrons and the bilepton mass
$M_Y$ were varied as free parameters to determine the $\chi^2$
distribution. The $95\%$ confidence level bound corresponds to an
increase of the $\chi^2$ by $5.99$ with respect to the minimum
$\chi^2_{min}$ of the distibution. Fig.9 presents the resulting
$95\%$ C. L. contour plots on the $(g_{3l},M_{Y})$ plane for the
nominal center-of-mass energies $\sqrt{s} = 500 GeV$ and $\sqrt{s}
= 1 \, TeV$. The unpolarized case is shown in Fig. 10. Since our
differential cross sections contain only even powers of $g_{3l}$,
it suffices to use positive values of the coupling in Figs. 9-10.
We also calculated the corresponding $95\%$ C. L. limits on the
bilepton mass at these NLC energies, considering only the
minimal $331$
model, in which a $Z^\prime$ exchange has to be taken into
account. The results are displayed in Table I.

\section{Conclusions}
Future linear colliders will provide an opportunity to look for
new particles and their interactions. In this paper we discussed
how to search for effects of a virtual vector bilepton in
polarized M\o ller scattering. Starting from a realistic
simulation of this process, we analyzed several
polarization-dependent observables that might provide strong
evidence of the existence of vector bileptons, and allow to
discriminate models or classes of models which predict these
particles, should any deviation from the standard expectations be
detected. We demonstrated that these polarization-dependent
observables are more sensitive to vector bileptons than their
unpolarized counterparts, in a large region of the allowed
parameter space. This sensitivity was found to be stronger at a
center-of-mass energy of $1 \, TeV$ than at $500 GeV$. The bounds on
the masses and couplings of the bileptons, derived from a $\chi^2$
estimator, indicate that it should be possible to probe mass
scales of up to several TeV for a signal of a vector bilepton.
Longitudinal polarization of the electron beams in M\o ller scattering
has proved useful to improve these bounds on the masses and couplings.
Although most of our calculations were carried out in the context of
the minimal $331$ model and an $SU(15)$ GUT model, we believe that our
overall conclusions could be extended to other models with bileptons.

\begin{table}
\caption{\label{table1}$95\%$ C. L. limits on the bilepton mass
in the minimal $331$ model, at NLC energies}
\begin{tabular}{lcr}
Polarization&$\sqrt{s}$=500 GeV&$\sqrt{s}$=1 \, TeV \\
\hline
unpolarized & 1230 GeV & 1815 GeV\\
polarized & 2529 GeV & 4574 GeV\\
\end{tabular}
\end{table}

\setcounter{figure}{0}
\begin{figure*}
\centerline{\epsfxsize=12cm\epsffile{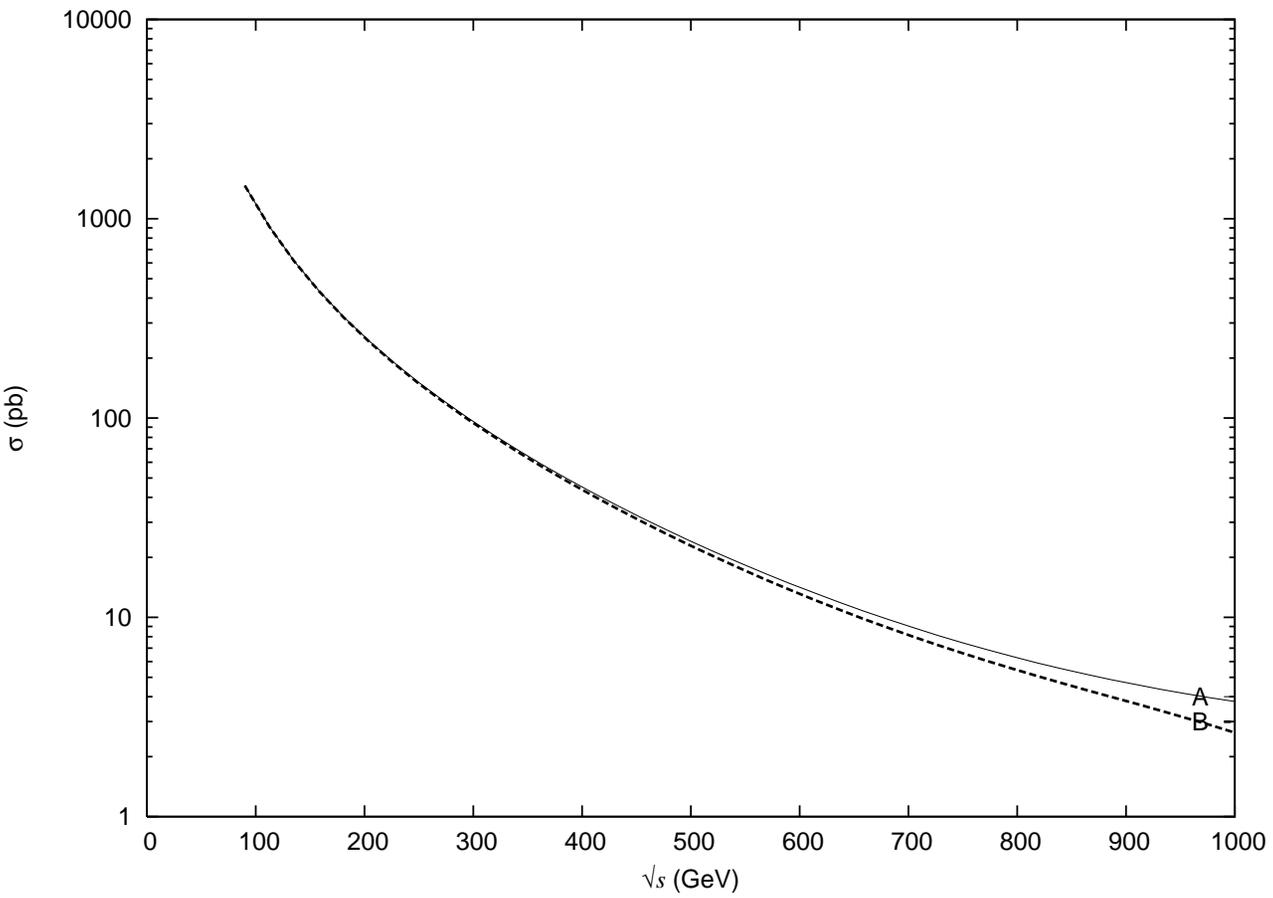}}
\caption{\label{fig1}{\rm Total cross section as a function of the
center-of-mass energy $\sqrt{s}$, for $M_Y = 1.2 \, TeV \,$ (B);
solid line (A) represents the standard model cross section.}}
\end{figure*}

\setcounter{figure}{1}
\begin{figure*}
\centerline{\epsfxsize=12cm\epsffile{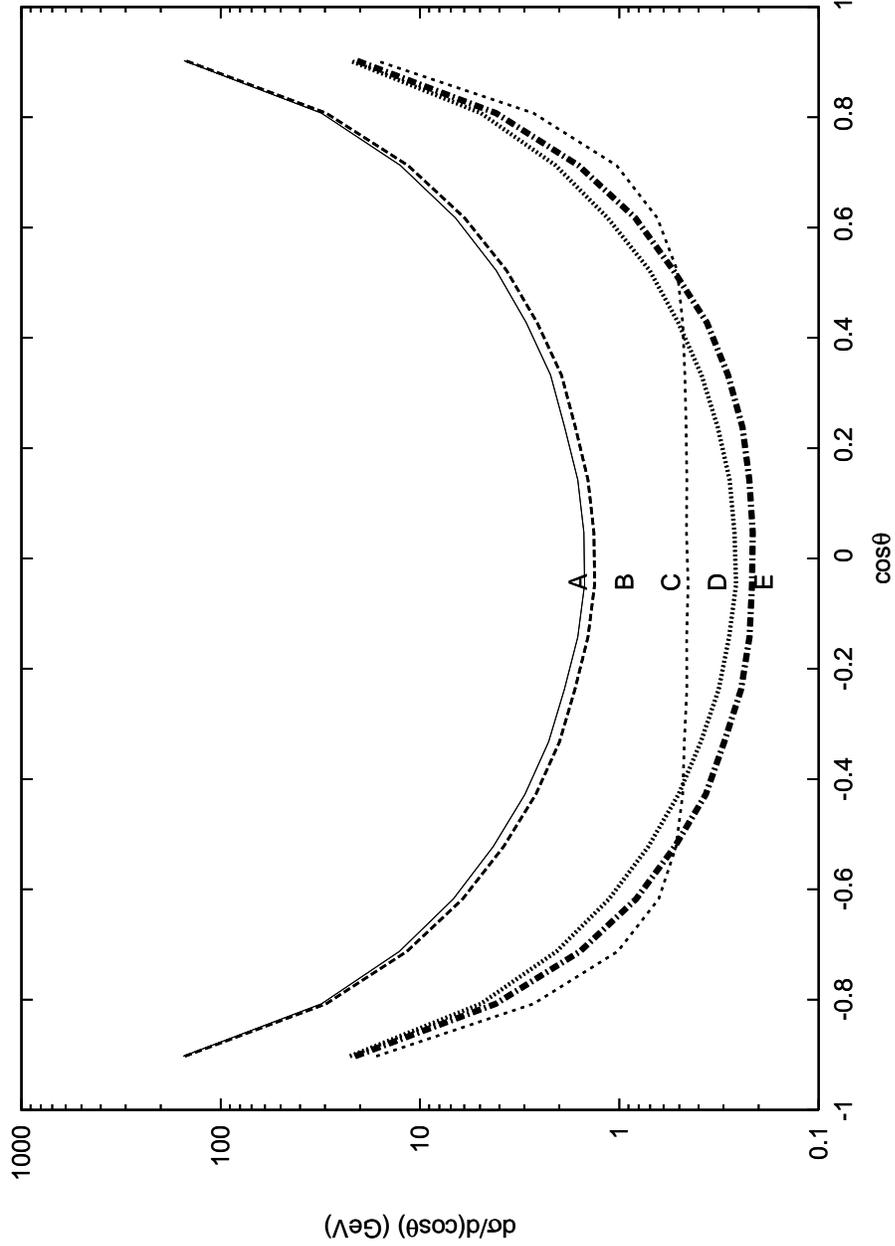}}
\caption{\label{fig2} Angular distribution of the final-state
electrons; Curves (A) and (D) show the angular spectra predicted
by the standard model at $\sqrt{s} = 500 GeV$ and $\sqrt{s} = 1 \,
TeV$ respectively, while (B) and (C) display the corresponding
angular distributions in the minimal {331} model. Curve (E) shows
the prediction for the SU(15) model at $\sqrt{s} = 1 \, TeV$.}
\end{figure*}

\setcounter{figure}{2}
\begin{figure*}
\centerline{\epsfxsize=12cm\epsffile{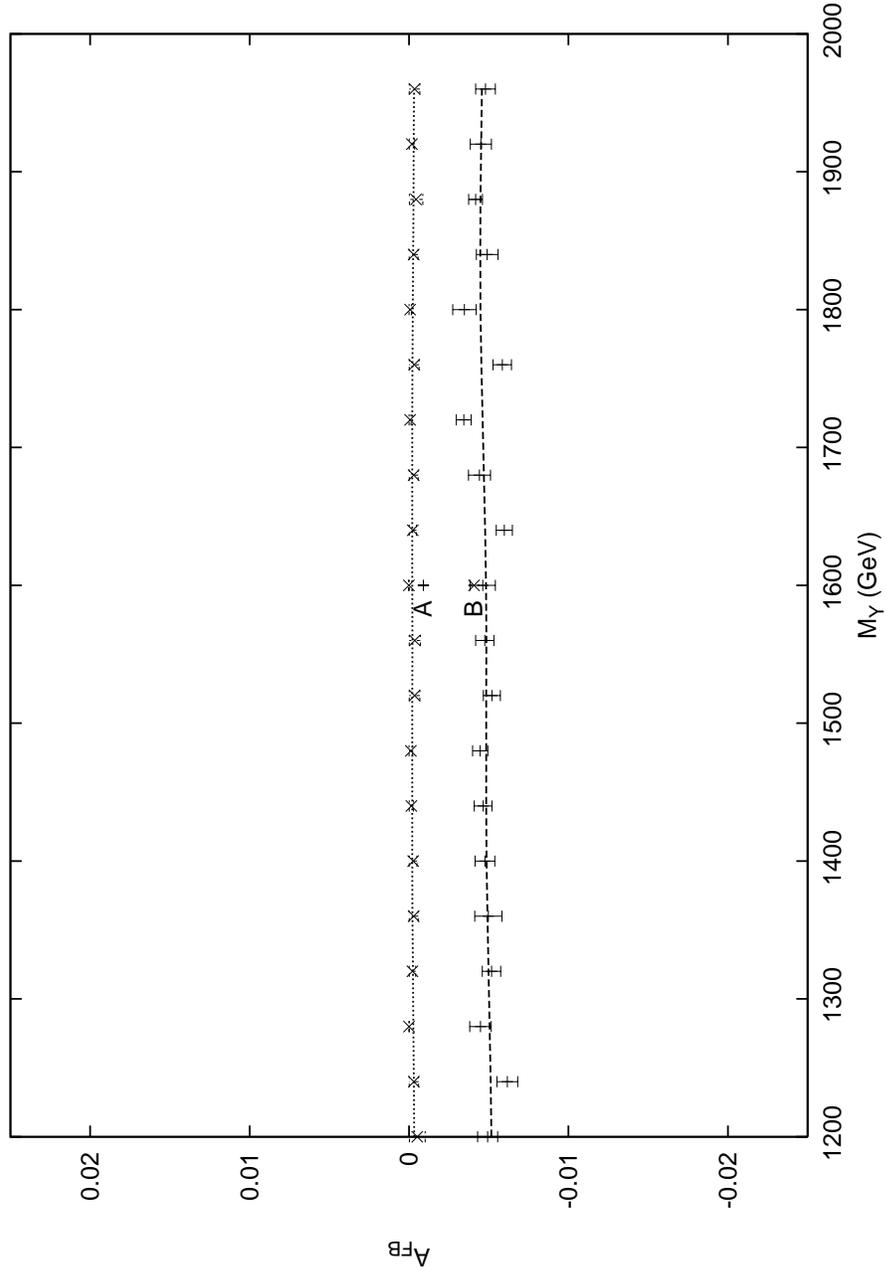}}
\caption{\label{fig3} Forward-backward asymmetry $A_{FB}$ for
several input masses $M_Y$, at $\sqrt{s} = 1 \, TeV$, according to
the minimal $331$ model (B). The upper curve (A) shows the
expected value for the standard model}
\end{figure*}

\setcounter{figure}{3}
\begin{figure*}
\centerline{\epsfxsize=12cm\epsffile{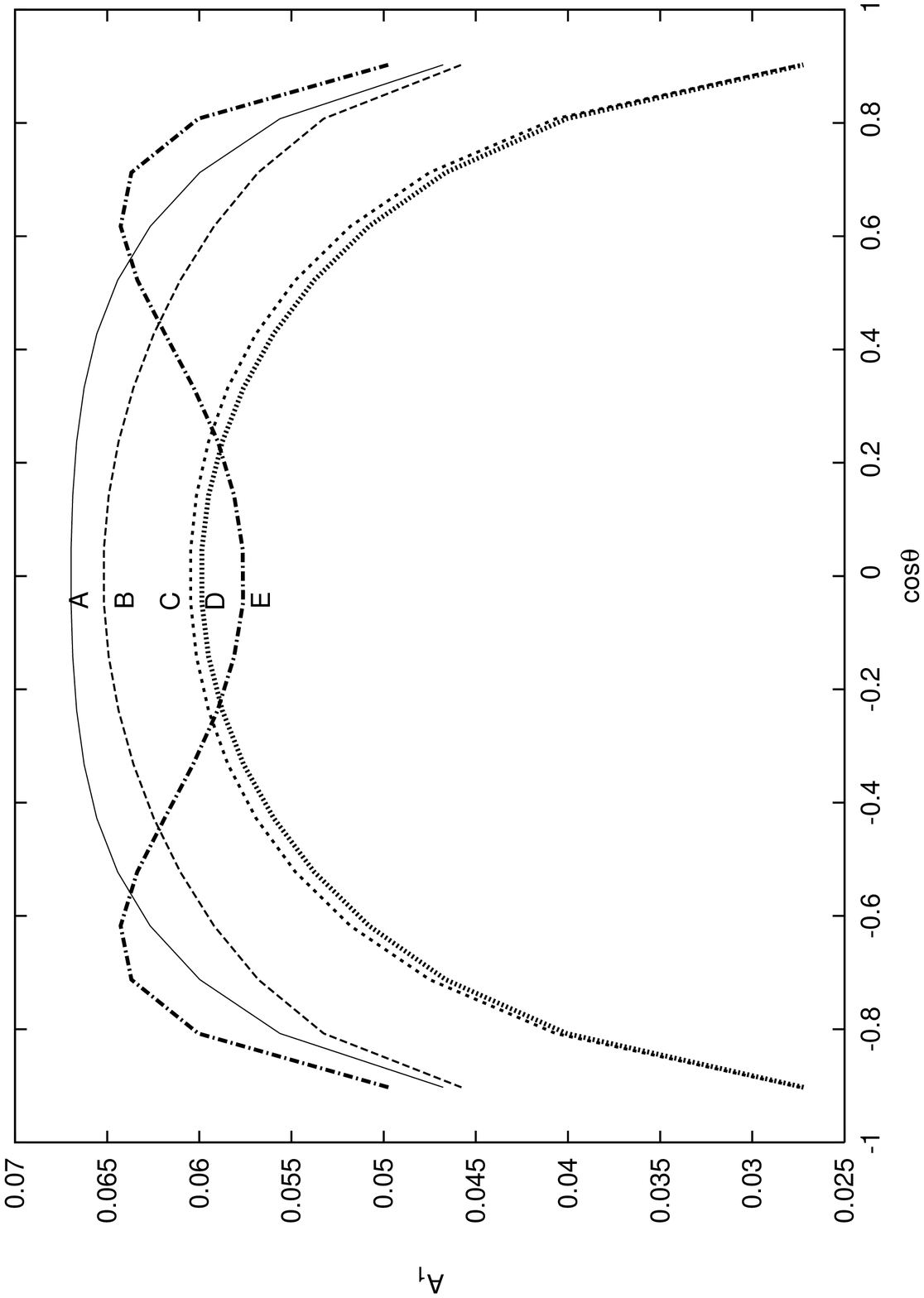}}
\caption{\label{fig4} Polar angle dependence of spin asymmetry
$A_1(cos\theta)$; Curve (A) shows the prediction for the SU(15)
model at $\sqrt{s} = 1 \, TeV$. Curves (D) and (B) show the
standard model predictions for $A_1(cos\theta)$ at $\sqrt{s} = 500
GeV$ and $\sqrt{s} = 1 \, TeV$ respectively, while (C) and (E)
represent the corresponding expectations for the minimal {331}
model, for a mass $M_Y = 1.2 \,  TeV$.}
\end{figure*}

\setcounter{figure}{4}
\begin{figure*}
\centerline{\epsfxsize=12cm\epsffile{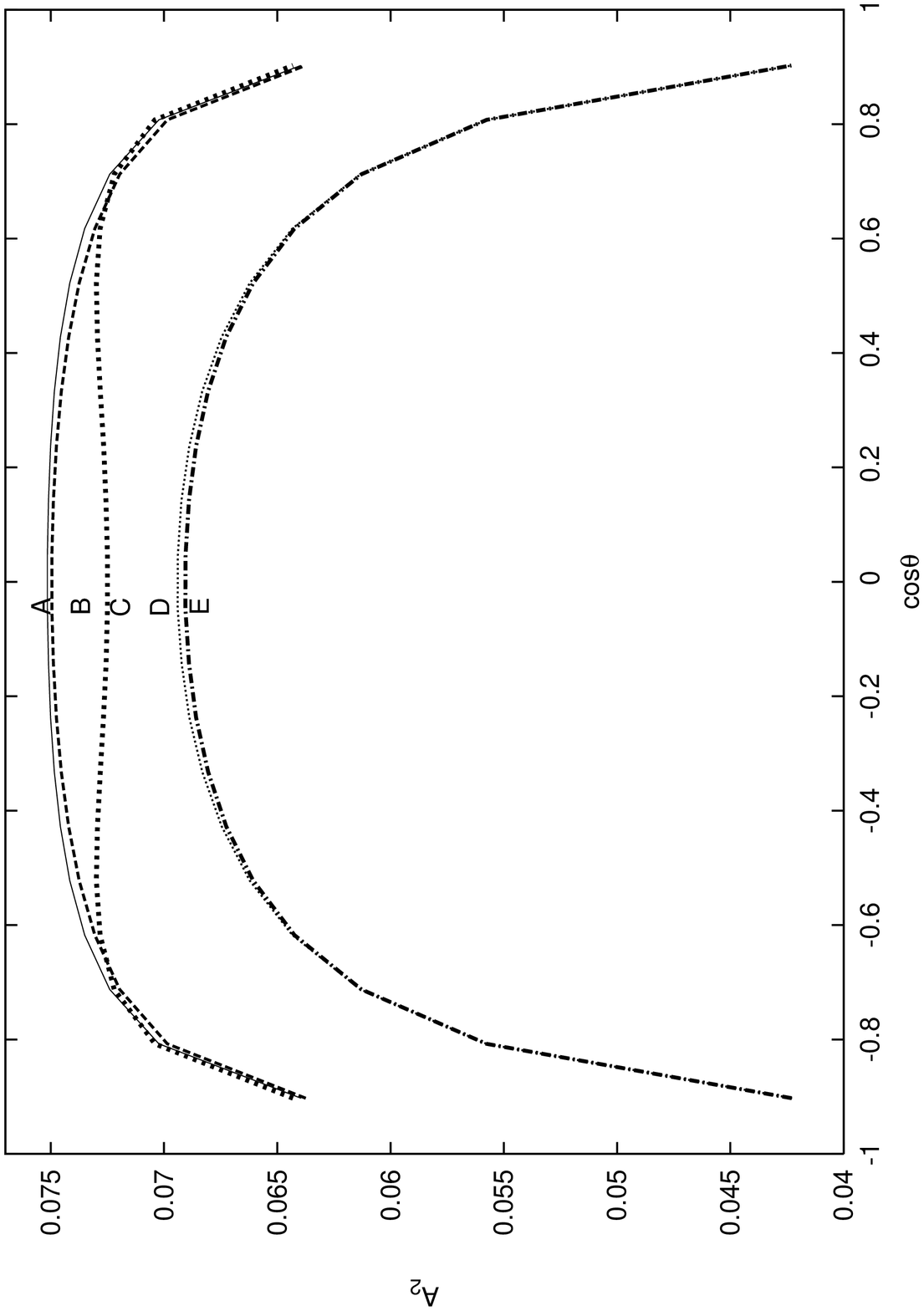}}
\caption{\label{fig5} Polar angle dependence of spin asymmetry
$A_2(cos\theta)$; Curve (A) shows the prediction for the SU(15)
model at $\sqrt{s} = 1 \, TeV$. Curves (B) and (D) show the
standard model predictions for $A_2(cos\theta)$ at $\sqrt{s} = 500
GeV$ and $\sqrt{s} = 1 \, TeV$ respectively, while (C) and (E)
represent the corresponding expectations for the minimal {331}
model, for a mass $M_Y = 1.2 \, TeV$. }
\end{figure*}

\setcounter{figure}{5}
\begin{figure*}
\centerline{\epsfxsize=12cm\epsffile{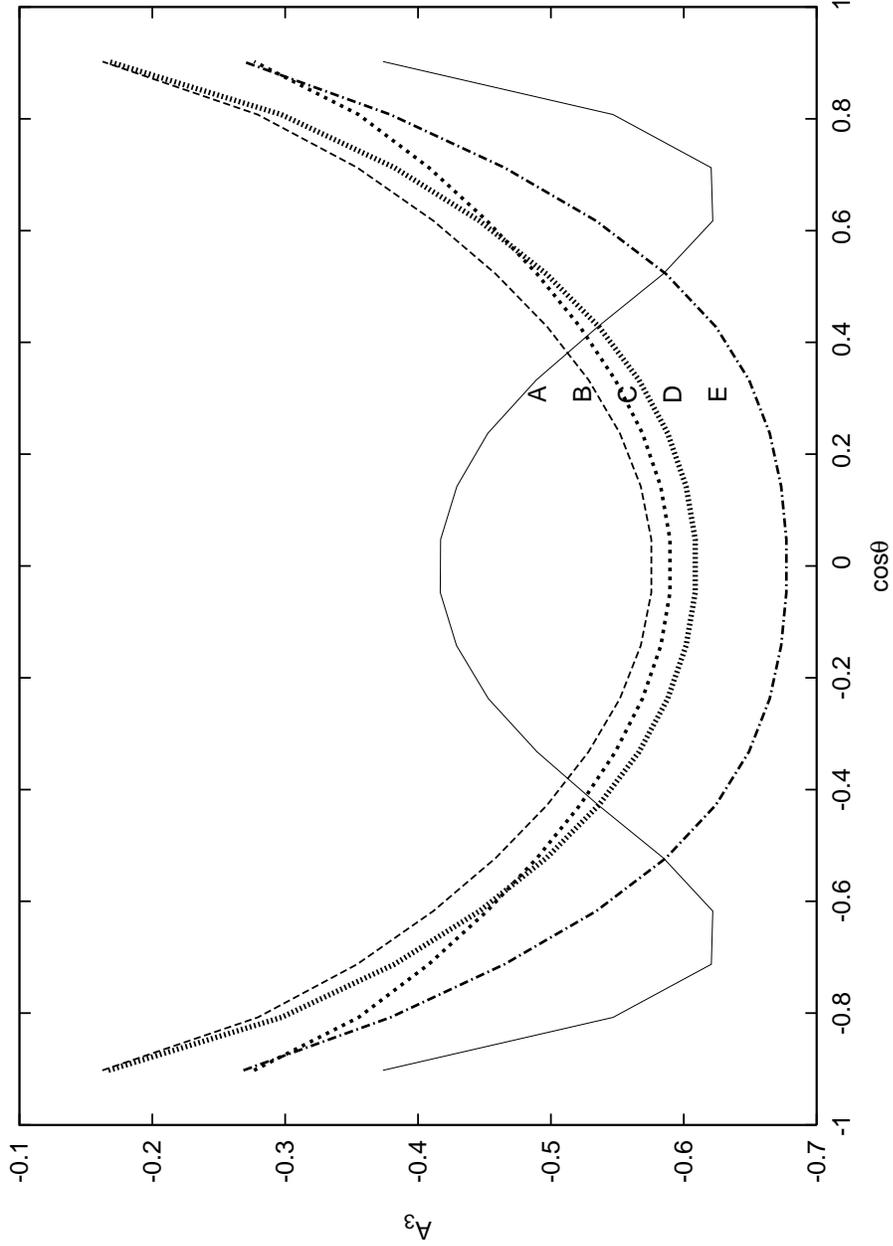}}
\caption{\label{fig6} Polar angle dependence of spin asymmetry
$A_3(cos\theta)$;  Curves (B) and (C) show the standard model
predictions for $A_3(cos\theta)$ at $\sqrt{s} = 500 GeV$ and
$\sqrt{s} = 1 \, TeV$ respectively, while (D) and (A) represent
the corresponding expectations for the minimal {331} model, for a
mass $M_Y = 1.2 \, TeV$. Curve (E) shows the prediction for the
SU(15) model at $\sqrt{s} = 1 \, TeV$.}
\end{figure*}

\setcounter{figure}{6}
\begin{figure*}
\centerline{\epsfxsize=12cm\epsffile{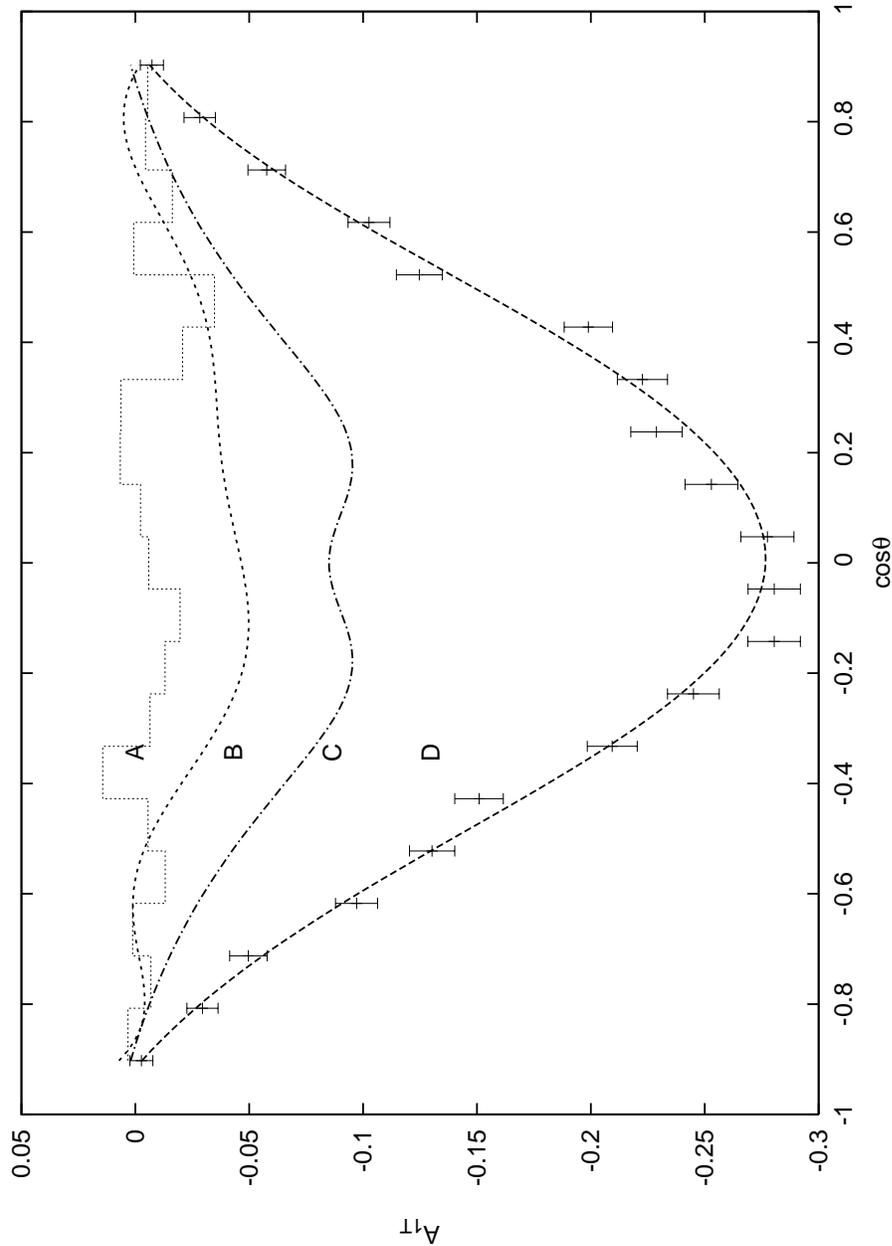}}
\caption{\label{fig7} Transverse polarization asymmetry
$A_{1T}(cos\theta)$ as a function of $cos\theta$, at $\sqrt{s} = 1
\, TeV$. The curve with error bars (D) corresponds to the minimal
$331$ model with $M_Y = 800 GeV$, whereas curve (C) corresponds to
the $SU(15)$ GUT model prediction for the same mass. Curve (B)
represents the behavior of the asymmetry for a $1.2 \, TeV$
bilepton in the minimal $331$ model, and the histogram (A)
corresponds to the standard model expectation. The error bars for
the upper curves are similar to those of the lower curve on a
bin-to-bin basis.}
\end{figure*}

\setcounter{figure}{7}
\begin{figure*}
\centerline{\epsfxsize=12cm\epsffile{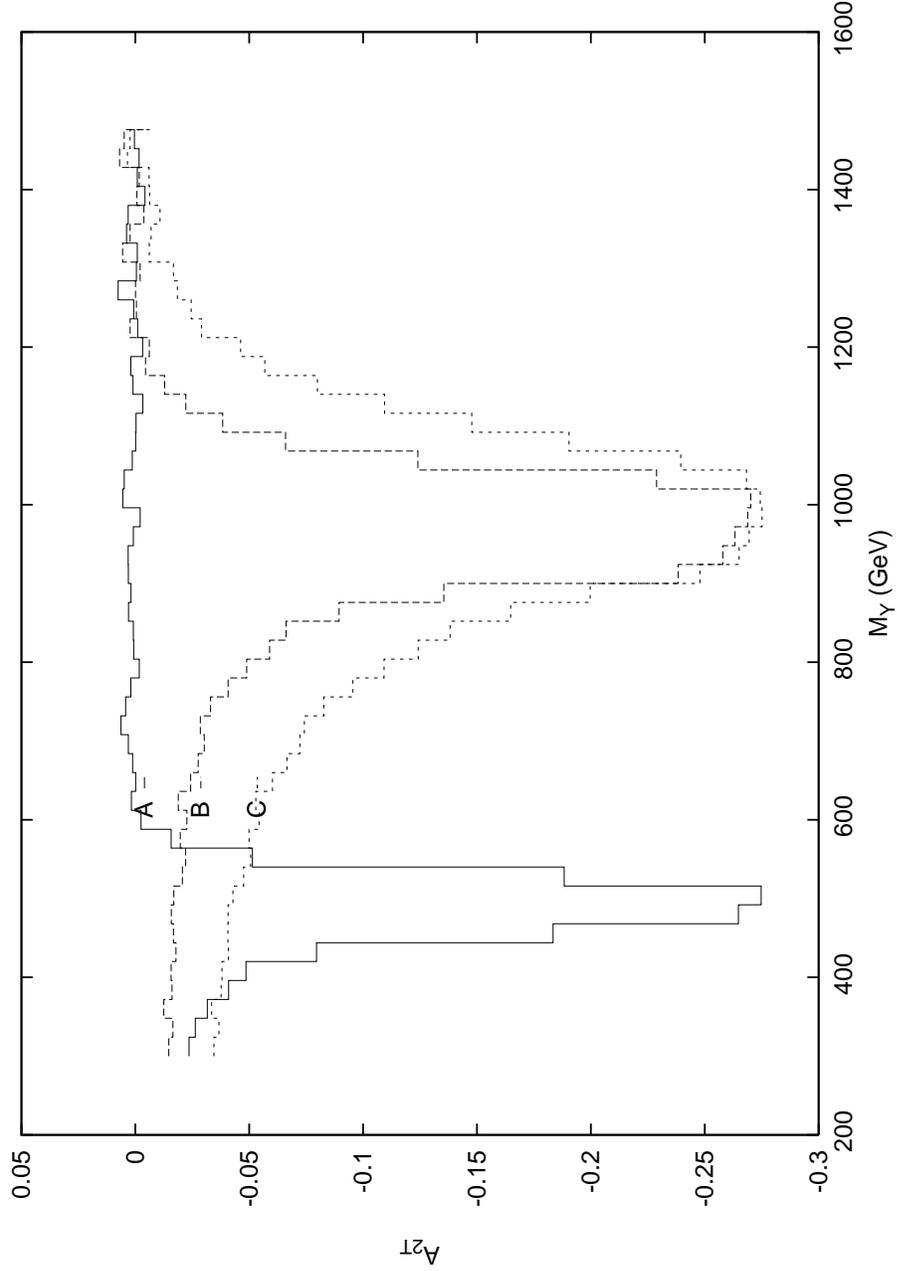}}
\caption{\label{fig8} Transverse polarization asymmetry $A_{2T}$
as a function of $M_Y$. The solid histogram (A) refers to the
minimal $331$ model prediction at an energy $\sqrt{s} = 500 GeV$
and the long-dashed histogram (B) to the $SU(15)$ model at $1 \,
TeV$. The resulting asymmetry $A_{2T}$ for the minimal $331$ model
at $\sqrt{s} = 1 \, TeV$ is is represented by the dashed histogram
(C).}
\end{figure*}

\setcounter{figure}{8}
\begin{figure*}
\centerline{\epsfxsize=12cm\epsffile{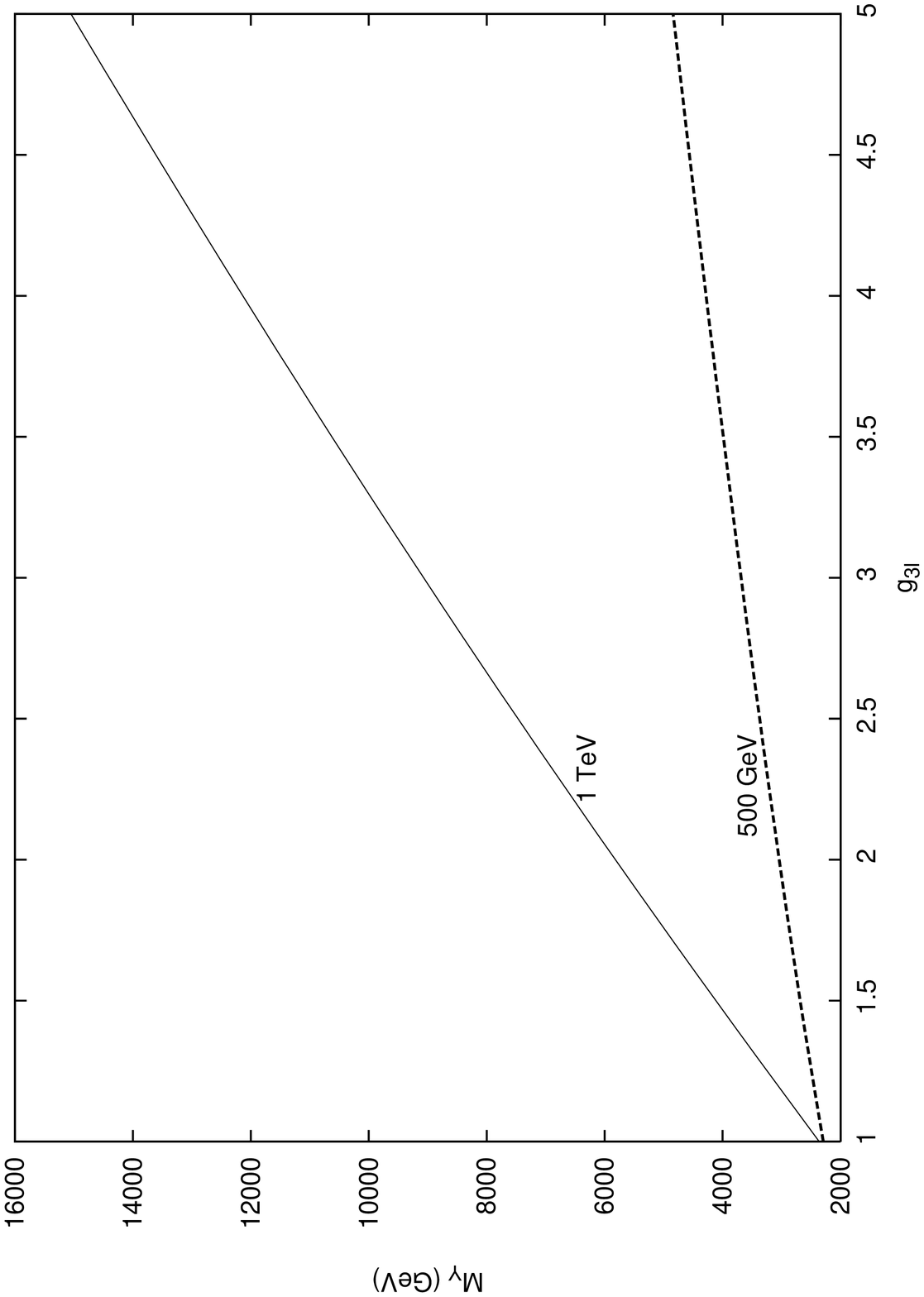}}
\caption{\label{fig9} $95\%$ C. L. contour plots on the
$(g_{3l},M_Y)$ plane for longitudinally polarized M\o ller
scattering, at the NLC center-of-mass energies $\sqrt{s} = 500
GeV$ (lower curve) and $\sqrt{s} = 1 \, TeV$ (upper curve). }
\end{figure*}

\setcounter{figure}{9}
\begin{figure*}
\centerline{\epsfxsize=12cm\epsffile{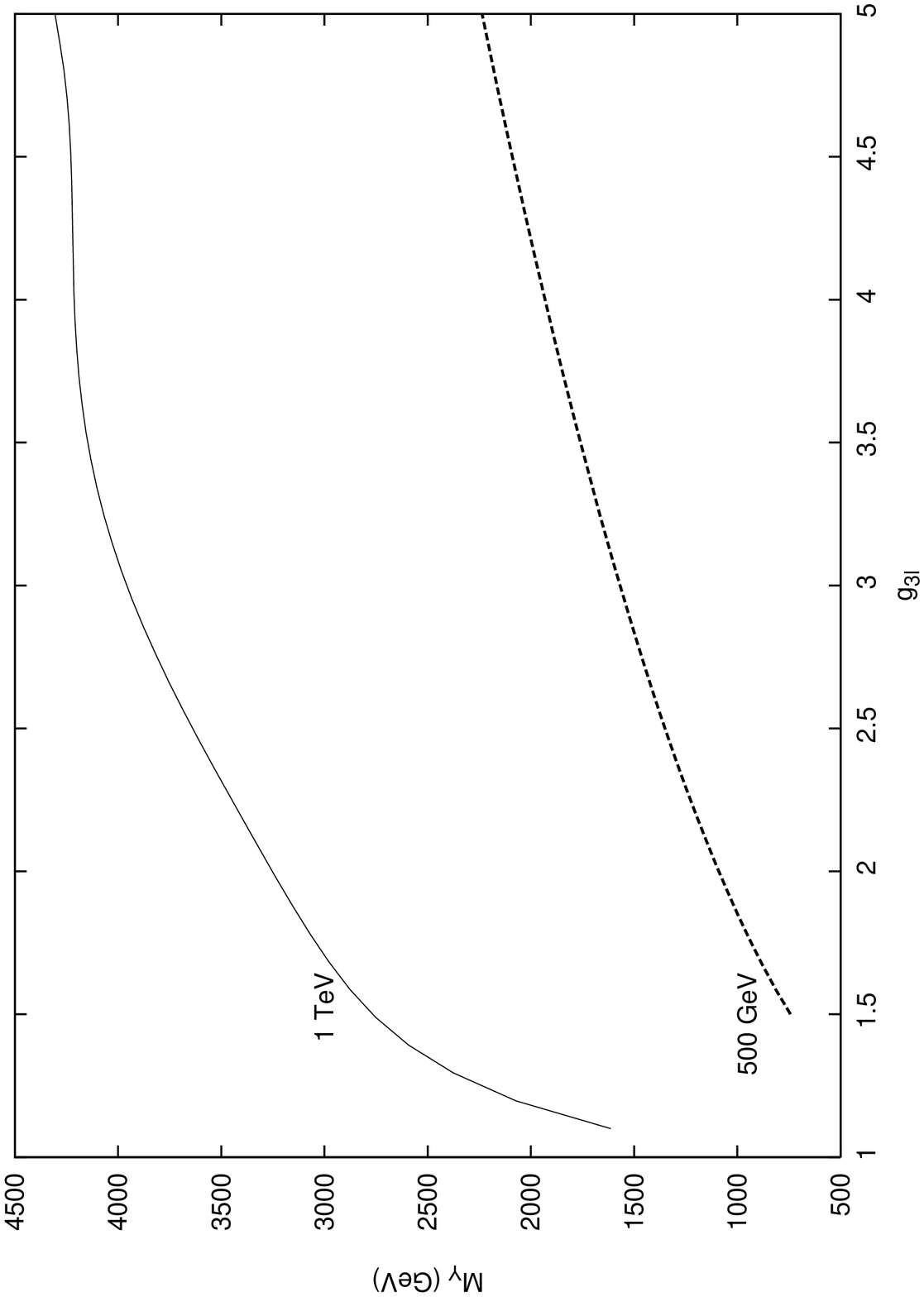}}
\caption{\label{fig10}$95\%$ C. L. contour plots on the
$(g_{3l},M_Y)$ plane for unpolarized M\o ller scattering, at the
NLC center-of-mass energies $\sqrt{s} = 500 GeV$ (lower curve) and
$\sqrt{s} = 1 \, TeV$ (upper curve).}
\end{figure*}

\end{document}